\documentclass[fleqn,usenatbib]{mnras}

\usepackage{newtxtext,newtxmath}

\usepackage[T1]{fontenc}

\DeclareRobustCommand{\VAN}[3]{#2}
\let\VANthebibliography\thebibliography
\def\thebibliography{\DeclareRobustCommand{\VAN}[3]{##3}\VANthebibliography}

\usepackage{graphicx}	
\usepackage{amsmath}	
\usepackage{bm}
\usepackage{subcaption}
\usepackage{siunitx}
\usepackage{xcolor}

\graphicspath{{./Figures/}}

\newcommand{\Pe}{\mathrm{Pe}}
\renewcommand{\Pr}{\mathrm{Pr}}
\newcommand{\Pm}{\mathrm{Pm}}

\defcitealias{Perrone2022}{PL22a}
\defcitealias{Perrone2022a}{PL22b}

\title[Adiabatic MTI]{Magneto-Thermal Instability in Galaxy Clusters -- III. The Limit of Adiabatic Stratification}

\author[L. M. Perrone et al.]{
	Lorenzo M. Perrone,$^{1}$
	Henrik Latter$^{2}$\thanks{E-mail: hl278@cam.ac.uk}
	\\
	$^{1}$Leibniz-Institut f\"{u}r Astrophysik Potsdam (AIP), An der Sternwarte 16, D-14482 Potsdam, Germany\\
	$^{2}$Department of Applied Mathematics and Theoretical Physics, University of Cambridge, Wilberforce Rd, Cambridge CB3 0WA, UK
}

\date{Accepted XXX. Received YYY; in original form ZZZ}

\pubyear{2026}

\begin{document}
\label{firstpage}
\pagerange{\pageref{firstpage}--\pageref{lastpage}}
\maketitle

\begin{abstract}
In the hot and dilute intracluster medium of galaxy clusters, large-scale buoyancy instabilities can develop due to the transport of heat along magnetic field lines. In particular, the peripheries of galaxy clusters are unstable to the magneto-thermal instability (MTI), which may contribute to the observed levels of turbulence. Recent theoretical and numerical work has revealed that the stable background entropy stratification controls the nonlinear saturation of the instability, by setting the strength and the integral scale of the resulting turbulent state. However, observations of the periphery of galaxy clusters show that the radial entropy profiles near the virial radii $R_{500}$ may be flatter than predicted by models of smooth gravitational accretion. 
This motivates us to investigate the saturation of the MTI in adiabatic (buoyantly neutral) atmospheres, using both phenomenological approaches and Boussinesq numerical simulations, carried out with the pseudospectral code {\sc SNOOPY}. We find that the adiabatic MTI saturates in a state characterised by the formation of large-scale plumes and their destruction by shear instability, yielding a new scaling law for the saturated turbulent kinetic energy, $\sim$$\chi \omega_T$, as the adiabatic limit is approached, where $\chi$ is the effective thermal diffusivity and $\omega_T$ is the MTI frequency. This predicts that the MTI plumes may achieve near sonic speeds in cluster outskirts, thus providing significant turbulent pressure support, even in the face of suppressed thermal conduction.     
\end{abstract}




\section{Introduction}

The outskirts of galaxy clusters are turbulent environments: continuous mass accretion, mergers, and infall of substructures are all expected to drive disordered motions on scales of hundreds of $\si{kpc}$  \citep{Schuecker2004,Ichinohe2015,Eckert2017a,Simionescu2017,Simionescu2019}. Separately, large-scale buoyancy instabilities, intrinsic to the plasma itself, can significantly stir up the plasma  \citep{Balbus2000,Parrish2008,McCourt2011,Parrish2012c,Perrone2022a}. 
In concert, these turbulent motions contribute appreciably to the pressure support in the periphery of galaxy clusters \citep[up to $10-30 \%$ of thermal pressure, see, e.g.,][]{Ettori2019,Eckert2019}. 

In the hot and dilute intra-cluster medium (ICM), charged particles rapidly gyrate around magnetic field lines but only infrequently scatter due to Coulomb collisions; thus, transport perpendicular to the field lines is inhibited, with heat transfer mostly confined along magnetic field lines \citep{Braginskii1965a}.
This property may help explain a variety of phenomena, such as the apparent suppression of heat conduction and the dynamical stability of cold fronts and bubbles \citep{Vikhlinin2001,Vikhlinin2002, Markevitch2007,ZuHone2014,ZuHone2016a,Richard-Laferriere2023}. 
Moreover, it renders inapplicable the classic Schwarzschild criterion for convection \citep[which is controlled by the sign of the entropy gradient;][]{Schwarzschild1906}. Rather, it is now the temperature gradient that controls the onset of instabilities \citep{Balbus2000}. In the periphery of galaxy clusters the background temperature decreases with radius, making the ICM unstable to the \textit{magneto-thermal} instability \citep[MTI;][]{Balbus2000,Balbus2001}.

In \citet[][hereafter \citetalias{Perrone2022} and \citetalias{Perrone2022a}]{Perrone2022,Perrone2022a}, we performed a comprehensive study of the MTI in two and three dimensions, both theoretically and numerically, using the pseudospectral code {\sc SNOOPY} in the Boussinesq approximation.
We showed that at saturation the MTI produces a state of buoyancy-driven
turbulence composed of density and velocity fluctuations over
a wide range of scales, sustained by energy drawn from the background temperature gradient. Importantly, the turbulent properties at saturation
follow clear power laws. In particular, we found that the stable background entropy stratification is critical for the nonlinear saturation of the instability, and controls the strength and the integral scale of the MTI-turbulence at saturation. Subsequent studies employing fully compressible MHD simulations confirmed the validity of the main scaling laws of MTI-turbulence \citep{Kempf2023a,Kempf2025}. The contribution of the MTI to the turbulent levels observed in the outskirts of galaxy clusters is still a subject of ongoing work. In particular, other competing sources of turbulence, such as gravitational accretion of substructures and mergers, may overwhelm the MTI. Simple models of the MTI that include external forcing to account for other sources of turbulence show that the MTI is rather resilient \citep{Perrone2024}, but more realistic studies in cosmological setups are necessary to draw firmer conclusions \citep{Talbot2024}.

\subsection{Entropy flattening in the periphery of galaxy clusters}

A number of observations \citep[e.g.][]{George2009,Simionescu2011,Urban2011,Walker2012,Simionescu2017} suggest that in several nearby galaxy clusters the entropy profiles close to the virial radius may be flatter than predicted by models of smooth gravitational accretion \citep{Tozzi2001,Voit2005c}. 
This discrepancy has been attributed to gas clumping in the ICM \citep{Simionescu2011,Nagai2011}, which would lead to an overestimation of the X-ray emission (dominated by the bigger clumps), naturally explaining the higher density and lower entropy measured in the outskirts. Correcting for the clumping generally results in a better agreement of the entropy profiles beyond the virial radius with expectations  \citep{Tchernin2016,Ghirardini2019,Zhu2021}. 
In several other clusters, however, even when the clumps are accounted for the entropy profiles remain flat or significantly lower than the baseline profile beyond $R_{500}$ \citep{Walker2013,Fusco-Femiano2014,Ghirardini2017}, which is the radius such that the mean density inside is 500 times the critical density. In some of these systems, a non-thermal pressure support of about $40\%$ of the total pressure beyond the virial radius may contribute to the observed flattening \citep{Ghirardini2017}. 
Other mechanisms to explain the entropy flattening have been proposed, among which we mention the possible difference in the temperatures between electrons and ions in the outskirts \citep{Hoshino2010,Akamatsu2011}, or a weakening of the accretion shock \citep{Lapi2011,Cavaliere2011}.

\subsection{Implications for the MTI}

The implications of a flatter entropy profile on the MTI are significant.
In the stably stratified plasma studied in \citetalias{Perrone2022,Perrone2022a}, the entropy stratification provides a `stabilizing' effect  in the form of an outer scale for the turbulence, which we call the buoyancy scale. At this scale, the restoring buoyancy forces balance the driving by the MTI. As the stratification is weakened, the buoyancy scale becomes increasingly larger, and the flow more turbulent. In fact, in the limit of very weak, or neutral, entropy stratification (hereafter referred to as `adiabatic-MTI') the scaling laws derived in our previous work formally diverge.

The issue partially stems from the absence of a physical (outer) length-scale (such as the pressure/density scale-height) in the Boussinesq approximation, which should intervene, replace the buoyancy scale, and thus circumvent the divergence. In its absence, two saturation scenarios emerge for adiabatic Boussinesq simulation: (a) the MTI saturates by reaching the box size, possibly producing domain-filling structures that resemble elevator modes, i.e., solutions which are characterised by predominantly vertical velocities with alternating sign in the horizontal direction that typically appear in Rayleigh-Benard convection with vertical periodic boundary conditions, \citep[see][]{Calzavarini2006,Lohse2024} or the so-called `open-field' MTI solutions \citep[see][]{Perrone2022}, which would mean the adiabatic saturation is truly global (and possibly compressible) and thus Boussinesq models unsuitable; or (b) the system finds a saturation path, independent of thermodynamics, that relies on an intrinsic self-limiting process, such as via parasitic instabilities, that imposes a new saturation scale (shorter than the box size). In the latter case, it may still be possible to obtain physical insights via Boussinesq simulations (which benefit from being simpler and more physically transparent) 
and then to extrapolate their results to the conditions found in the periphery of galaxy clusters. 

\subsection{Theoretical approach and paper plan}

In this paper, we take various approaches to understand MTI saturation in the adiabatic limit and, in particular, generalise the \citetalias{Perrone2022} scaling law for the saturated turbulent kinetic energy. Our main goal is to assess which of the two scenarios, sketched above, is realized in practice.
 
We begin, in Section~\ref{sec:mti-saturation-adiabatic}, by offering phenomenological arguments for MTI saturation, based on balancing the large-scale MTI structures/plumes the adiabatic limit favours against the parasitic shear instability that would limit their sizes and amplitudes. We then provide a set of semi-analytical and numerical calculations to illustrate, and indeed confirm, the general behaviour and scaling laws these arguments predict. Section~\ref{sec:theoretical-approach} briefly presents the governing equation of our model, while Section~\ref{sec:1d-model} supplies semi-analytical nonlinear solutions, representing large-scale runaway MTI `plumes' (or `elevator flows'), that (in the absence of viscosity) grow without bound in the absence of stable stratification. These archetypes provide useful references to help identify and understand large-scale structures in our nonlinear 3D simulations of Section~\ref{sec:numerical-experiments}. There we find the saturation of the adiabatic MTI controlled by the emergence of such plumes and their destruction by shear instability, and undertake a small parameter sweep that yields a new scaling law for the kinetic energy, independent of global considerations or compressibility. 
In Sections~\ref{sec:discussion} and~\ref{sec:conclusions}  we discuss the implications of these results.

\section{MTI saturation in adiabatically stratified clusters}\label{sec:mti-saturation-adiabatic}

Before exhibiting specific numerical calculations, we provide a phenomenological account of MTI saturation in an adiabatic atmosphere. Thereby, we construct upper bounds and possible scaling laws for the saturated turbulent kinetic energy.

To get started we define some necessary quantities. First, we denote by $\omega_T$ the MTI frequency, which corresponds to the maximum growth rate of the linear instability (optimised over wavenumbers). Meanwhile $N$ is the Brunt-V\"ais\"al\"a frequency, which determines the buoyant response of the system in a stably-stratified environment, in the absence of thermal conduction. They are defined by
\begin{align}\label{eq:freq}
	\omega_T^2 =  \frac{1}{\rho}\frac{\partial  p}{\partial R}  \frac{\partial \ln T}{\partial R} , \quad
	N^2 = - \frac{1}{\gamma \rho}\frac{\partial  p}{\partial R}  \frac{\partial \ln (p \rho^{-\gamma})}{\partial R} ,
\end{align}
where $\rho$, $p$, $T$, $\gamma$, and $R$ are density, pressure, temperature, adiabatic index, and spherical radius. In addition, we take the thermal diffusivity of the plasma to be $\chi$.

Next, we recall the Boussinesq MTI saturation of \citetalias{Perrone2022} and \citetalias{Perrone2022a}, which reveals that the saturated specific kinetic energy $K$ and integral scale $\ell_i$ obey 
 \begin{equation} \label{scaling}
 K\sim \chi\omega^3_T/N^2, \qquad \ell_i\sim \sqrt{\chi\omega_T}/N.
 \end{equation}
Finally, the linear MTI modes achieve their maximum growth on scales less than the diffusive length $\ell_{\chi}= \sqrt{\chi/\omega_T}$. 

\subsection{An upper bound on the saturated kinetic energy}

Consider a model of a global cluster periphery with $N > 0$, in which the MTI has saturated on an integral scale $\ell_i\ll H$, where $H$ is the pressure scale-height of the plasma. Suppose we slowly decrease $N$. As a result, the integral scale will slowly lengthen, in accordance with the scaling law detailed above. At some point, however, we will reach the regime $\ell_i\sim H$, and no matter how much lower $N$ goes the integral scale can no longer increase: simply, the cluster is insufficiently large to contain longer turbulent scales. We denote by $N_{\mathrm{c}}$ the critical buoyancy frequency that yields $\ell_i\sim H$ and posit that values of $0 \leq N<N_{\mathrm{c}}$ (including $N=0$) will yield an MTI saturation equivalent to the $N=N_{\mathrm{c}}$ case.

To obtain an estimate for $N_{\mathrm{c}}$ we use Eq.~\eqref{scaling} and rework $\ell_i\sim H$ to find 
$N_{\mathrm{c}} \sim (\ell_{\chi}/H) \omega_T$. To obtain the (specific) kinetic energy at this critical value, we insert $N_{\mathrm{c}}$ into our scaling law for $K$, to obtain the critical saturated kinetic energy
\begin{align}\label{estimate}
    K_{\mathrm{c}} \sim  H^2 \omega_T^2 \sim \left(\frac{d \ln T}{d \ln \rho}\right) c_{\mathrm{s}}^2 < c_{\mathrm{s}}^2,
\end{align}
where $c_{\mathrm{s}}$ is the sound speed of the gas, and the last inequality follows from assuming $N=0$. Thus, saturated velocities, in this case, are at most transonic.
These estimates also follow from dimensional arguments: Suppose the saturated global adiabatic MTI produces plumes (or cells) of a maximum size $\sim H$. The characteristic turbulent MTI timescale cannot exceed $\omega_T$. We might then expect the maximum rms flow in these global plumes to be $u_{\mathrm{rms}} < H \omega_T$.

We stress that the estimate in Eq.~\eqref{estimate} is only an upper bound on $K$ in an adiabatic atmosphere. It assumes that the local Boussinesq scalings apply to (compressible) structures on global lengthscales; in fact, physical effects neglected by this approximation may intervene to limit MTI motions. Because the predicted turbulent speeds are subsonic, compressible effects (such as shock dissipation) are unlikely to intervene critically. On the other hand, if the MTI saturation is dominated by a network of large-scale plumes, consisting of updrafts and downdrafts \citep[cf.\ solar convection; e.g.,][]{Nordlund}, then parasitic modes feeding off their velocity shear could limit the MTI motions before they reach the pressure scale-height. We explore this idea through the rest of this section.   

\subsection{Saturation of large-scale MTI plumes by parasitic modes}

As shown by the direct numerical simulations of Section~\ref{sec:numerical-experiments}, we indeed do find that the adiabatic MTI develops large-scale plumes, as described above. While Eq.~\eqref{estimate} gives an an upper bound on their kinetic energy at saturation, this bound will not be reached in practice if the flows are attacked by parasitic KHI, feeding on the horizontal velocity shear associated with the updrafts and downdrafts. We hence posit a balance between the continuous production of plumes by the MTI and subsequent destruction by efficient KHI, a saturation that may take a cyclic form over time. This leads us to the following scaling for the saturated kinetic energy:
\begin{align} \label{Bestimate}
    K \sim \ell_{\chi}^2 \omega_T^2 .
\end{align}
The scaling can be justified on physical and dimensional grounds, noting that the only lengthscales in the problem are $\ell_\chi$ and $H$, and the only timescale is $1/\omega_T$. If $\ell_\chi \ll H$, and the KHI is efficient, then the MTI eddies (and associated KHI) remain near the MTI injection scale ($\sim \ell_\chi$), and thus the saturation remains `local' and independent of $H$, giving the above result. But if $l_\chi \sim H$, the two lengthscales are degenerate, so Eq.\eqref{Bestimate} also holds, and we recover the scaling for $K_\mathrm{c}$. 

Combining Eqs. \eqref{estimate} and \eqref{Bestimate} constrains the kinetic energy to lie in the following range: 
\begin{align}\label{eq:range_mti_saturation}
    \ell_{\chi}^2\omega_T^2 \lesssim K \lesssim H^2\omega_T^2 < c_{\mathrm{s}}^2,
\end{align}
which is relatively narrow once outside the Boussinesq regime. 

More generally, $K=g(\zeta)\ell_{\chi}^2 \omega_T^2 $, where $\zeta=\ell_{\chi}/H$, with $g(\zeta)$ a function potentially to be  determined from numerical simulations. (If there is a significant background magnetic field, viscosity, and/or resistivity, then $g$ will possess additional dependencies.) For $\zeta\to 0$ (the Boussinesq limit) and $\zeta\to 1$ (the global limit), we expect $g$ to approach two different order-one constants, which could enter \eqref{eq:range_mti_saturation} and sharpen the inequalities there.

\section{Theoretical approach and model}\label{sec:theoretical-approach}

Now that we have sketched out our theoretical expectations, from essentially physical arguments, we undertake several calculations to better illustrate, and ultimately validate, our picture of saturation. In this section, we briefly present the physical model, and its governing equations, used in the following two sections.

We work within the Boussinesq approximation, which is suited to describe the evolution of a small block of plasma in Cartesian coordinates in the presence of background gradient in both entropy and temperature. We imagine our box located at a given spherical radius $R_0$, with the vertical direction aligned with the radius, and characterised by a mean mass density $\rho_0$. We then assume that perturbed quantities vary on scales much shorter than the typical pressure (or density) scale-height of the system $H$, that their relative thermodynamic fluctuations are small, and that the turbulence velocities are subsonic \citep[see e.g.,][]{Spiegel1960}. In the Boussinesq approximation the background pressure gradient balances the local gravitational acceleration to leading order. As a result, in the momentum equation the gravitational force is replaced at next order by the buoyancy force, which is proportional to the density perturbation.
 
The Boussinesq MHD equations, including anisotropic thermal conduction, read \citepalias{Perrone2022}\footnote{In Eqs.~\eqref{eq:div_eq}--\eqref{eq:buoyancy_eq}, we have rescaled the magnetic field by a factor of $1/\sqrt{4 \pi \rho_0}$ so that it has the dimensions of a velocity, and the thermal diffusivity $\chi$ (in cgs units) absorbs a factor of $(\gamma -1)/\gamma$.}
\begin{gather}
	\nabla \cdot \bm u = \nabla \cdot \bm B = 0 ,\label{eq:div_eq} \\
	\left( \partial_t + \bm u \cdot \nabla \right) \bm u = - \frac{\nabla p_\text{tot} }{\rho_0} - \theta \bm e_z + \left( \bm B \cdot \nabla \right) \bm B + \nu \nabla^2 \bm u , \label{eq:mom_eq} \\
	\left( \partial_t + \bm u \cdot \nabla \right) \bm B = \left( \bm B \cdot \nabla \right) \bm u + \eta \nabla^2 \bm B, \label{eq:bfield_eq} \\
	\left( \partial_t + \bm u \cdot \nabla \right) \theta = N^2 u_z +  \nabla \cdot \left[ \chi \bm b \left( \bm b \cdot \nabla \right)\theta \right]	+  \omega_T^2 \nabla \cdot \left( \chi \bm b b_z\right), \label{eq:buoyancy_eq}
\end{gather}
where $\nabla p_\text{tot}$ is the gradient (vertical and horizontal) of the perturbed pressure including both thermal and magnetic pressure;
$\nu$ and $\eta$ are the fluid viscosity and magnetic resistivity; and $\theta = g_0 \delta \rho /\rho_0$ is the buoyancy variable, proportional to  the fractional density perturbation and to the local gravitational acceleration, $g_0$. The magnetic field has been rescaled by $\sqrt{4 \pi \rho_0}$ so that it has units of a velocity (Alfv\'{e}n velocity).
In what follows we assume $N^2 = 0$.

We solve these equations numerically in a finite domain of some size $L$ and then define the governing dimensionless parameters of the system: the Peclet number ($\Pe$), the Prandtl number ($\Pr$), and the magnetic Prandtl number ($\Pm$), are given by
\begin{equation}
\Pe= \frac{L^2\omega_T}{\chi}, \qquad \Pr=\frac{\nu}{\chi}, \qquad \Pm=\frac{\nu}{\eta}.
\end{equation}
It also useful to introduce the following specific `energies' (in units of velocity squared):
\begin{equation}
K=\frac{1}{2} \langle \bm{u}^2 \rangle, \quad M=\frac{1}{2}\langle\bm{B}^2 \rangle, \quad U=\frac{1}{2} \frac{\langle \theta^2 \rangle}{\omega_T^2} ,
\end{equation}
and corresponding root-mean-square values:
\begin{align}
    u_{\mathrm{rms}} = \langle\bm{u}^2 \rangle ^{1/2}, \quad B_{\mathrm{rms}} = \langle\bm{B}^2 \rangle ^{1/2}, \quad \theta_{\mathrm{rms}} = \langle \theta^2 \rangle ^{1/2},
\end{align}
where the angle brackets indicate an average over the numerical domain.  Occasionally, we calculate the energy in only one velocity or magnetic component, denoting such a quantity with the relevant subscript. Thus $M_z$ represents $\frac{1}{2}\langle B_z^2 \rangle$, for example. 

\section{A simple 1D model for large-scale plumes: elevator flows}\label{sec:1d-model}

In this section we show that without entropy stratification the Boussinesq equations admit a family of one-dimensional nonlinear solutions that exhibit transient or secular growth in the viscous or inviscid limit, respectively. These solutions, which we term `elevator flows', are characterized by the suppression of both anisotropic injection and dissipation in the buoyancy equation, and they can be reached asymptotically at the end of the exponential growth of an MTI eigenmode, thus prolonging its growth to nonlinear amplitudes.
They are important because they provide a local approximation to the large-scale plumes we expect to form in the global manifestation of the adiabatic MTI. Moreover, as we show later, these solutions commonly recur in fully-turbulent adiabatic MTI simulations, leading to cycles of elevator formation and subsequent breakdown by parasitic Kelvin-Helmholtz instabilities.

\subsection{Model equations}
\label{sec:1d-model-equations}

To calculate the form of these elevator solutions, we assume no background magnetic field, that all quantities depend only on the horizontal coordinate $x$, and that the solutions for the velocity and magnetic field comprise the vertical components, solely:
\begin{align}
	\bm u = u(x,t) \bm e_z, \qquad
	\bm B = B(x,t) \bm e_z, \qquad
	\theta = \theta(x,t).
\end{align}
Under these assumptions Eqs.~\eqref{eq:div_eq}-\eqref{eq:buoyancy_eq} reduce to:
\begin{subequations}
\label{eq:1Deqns}
\begin{align*}
	\frac{\partial u}{\partial t} = - \theta + \nu \frac{ \partial^2 u }{\partial x^2}, \hspace{0.2em} \mathrm{(14a)} \quad
	\frac{\partial B}{\partial t} =  \eta \frac{ \partial^2 B }{\partial x^2}, \hspace{0.2em} \mathrm{(14b)}\quad
	\frac{\partial \theta}{\partial t} = 0 . \hspace{0.2em} \mathrm{(14c)} 
\end{align*}
\end{subequations}
This family of solutions features a steady (`frozen-in') $\theta$, and the decoupling of vertical velocity from magnetic fields. Without placing any constraints on the initial profiles of $u$, $B$, and $\theta$ (except that $\partial_x \theta \neq 0$), we can see that the shear $\partial_x u$ will grow linearly in time (absent viscous damping), ultimately leading to destabilization by parasitic KHI. From a physical point of view, a `frozen-in' buoyancy perturbation acts like a constant forcing/acceleration. If $N^2>0$, a drag term would be present in the buoyancy equation, which would neutralize any growth via the excitation of g-modes.

\subsection{Explicit solution}\label{sec:1d-solution}

In the absence of viscosity, the momentum balance gives $u\propto \theta t$, otherwise it yields a forced 1D diffusion equation for $u$, which can be solved by classical techniques. In the simple but illustrative case of a periodic domain in $x$, choosing units so that the size of the domain is 1 and $u(x)=0$ at $t=0$, we set $u(x,t) =\sum u_n(t) \text{e}^{\text{i}n x}$ and $\theta (x)=\sum \theta_n\text{e}^{\text{i}n x}$, for unknown functions $u_n$ and known constants $\theta_n$, where $n$ is the mode wavenumber, with the additional constraint that $\theta_0=0$ (the density fluctuations have zero mean). It is straightforward to show that for $n\neq 0$,
\begin{equation}\label{eq:explicit_solution_fourier}
    u_n (t)= \frac{\theta_n}{\nu n^2}\left(\text{e}^{-\nu n^2 t}-1\right).
\end{equation}
At early times ($t \ll 1/ \nu n^2$), Taylor-expanding Eq.~\eqref{eq:explicit_solution_fourier} reveals that each Fourier coefficient $u_n (t)$ experiences linear growth. For small viscosities and low wavenumbers this period of algebraic growth can be considerable, leading to $u  \sim \theta /\nu$. At late times, growth of the Fourier coefficients $u_n$ peters out starting with modes with large $n$, and $u (x,t)$ approaches a time-independent function which can be found by integrating Eq.~(\ref{eq:1Deqns}a) directly, yielding $\nu^{-1}\int^x \left( \int^\xi \theta(\zeta) d\zeta \right)d\xi$.  (An analogous calculation, using Fourier transforms, shows the same behaviour for an unbounded domain.)  

\subsection{Relation to stably stratified open-field solutions}

The elevator flow solution just described shares some important aspects with the open-field solution studied in \citetalias{Perrone2022}, which was derived in the stably stratified limit ($N>0$) and with a uniform horizontal background magnetic field. 

The first point of difference concerns their saturation. If $N>0$ the growth of the open-field solution can be halted at sufficiently large perturbation amplitudes, and settles on g-mode oscillations. This is not possible for elevator flows, whose runaway growth can only be stopped by KHI (which is outside of the 1D model) or by viscosity.

The second point is in the nature of their early growth. Because of the background horizontal field, at its initial low amplitudes, the open-field solution coincides with  the linear MTI, and thus grows exponentially. The above elevator flow, owing to the absence of the background field, grows linearly with time and is driven by whatever profile of $\theta$ exists. 
However, the two phases of growth can be connected: Suppose $N=0$ but there is a weak background horizontal field; after the initial exponential growth of the linear MTI to sufficiently large amplitude, the perturbed magnetic field swamps the background field, and thus the destabilising linear and nonlinear heat flux terms disappear \citep[i.e. $h\gg 1$ in the notation of][]{Perrone2022}. The problem then resembles that of Eqs. \eqref{eq:1Deqns} but with $\theta$ taken as the end point of the linear growth of the eigenmode. The system then `locks onto' the linearly growing elevator solution (and not the unavailable g-mode solution).

\begin{figure}
    \centering
    \includegraphics[width=1.0\linewidth]{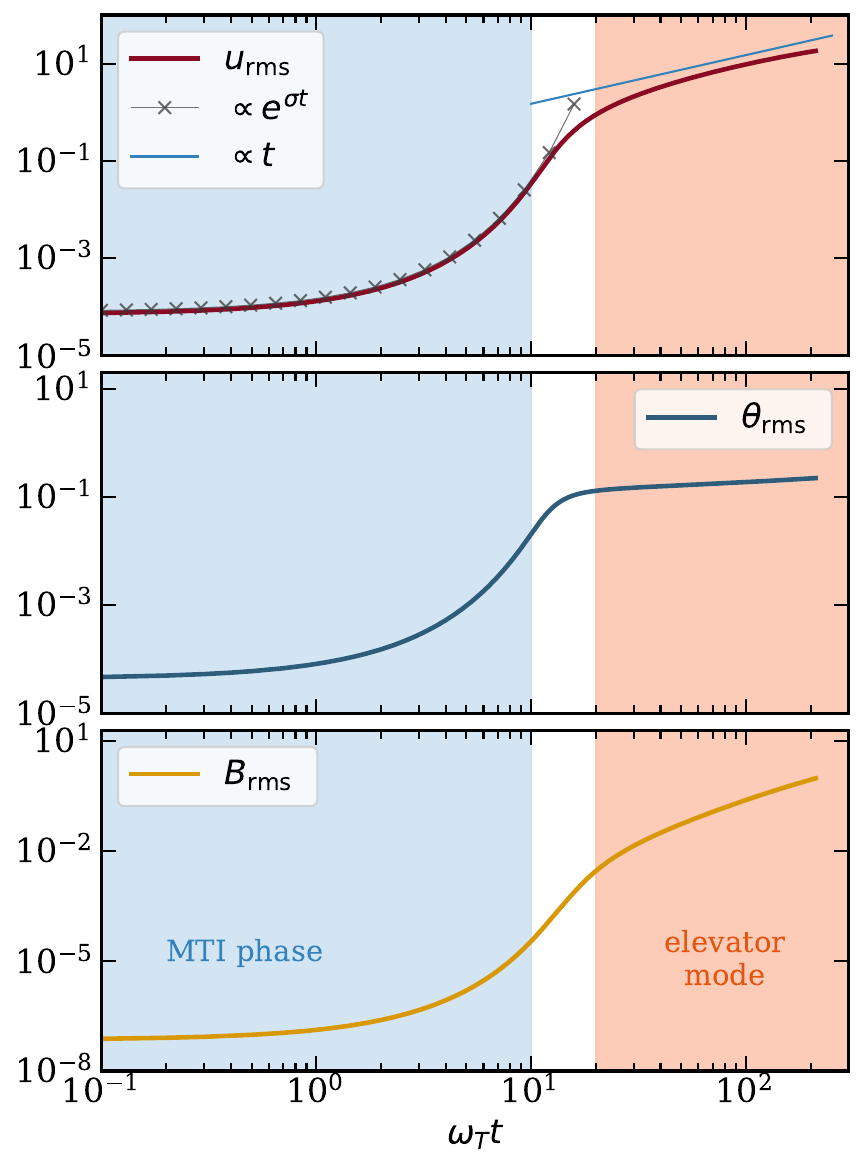}
	\caption{Root-mean-square values of velocity, buoyancy variable and magnetic field as a function of time in a 1D simulation of the adiabatic MTI with a weak background horizontal field. Around $t=10 \omega_T^{-1}$ one can observe the clear handover from the exponential growth of the MTI linear phase ($u_{\mathrm{rms}} \propto e^{\sigma t}$) to the elevator mode ($u_{\mathrm{rms}} \propto t$). }
	\label{fig:kin_pot_mag_energy}
\end{figure}

\begin{figure}
    \centering
    \includegraphics[width=1.0\linewidth]{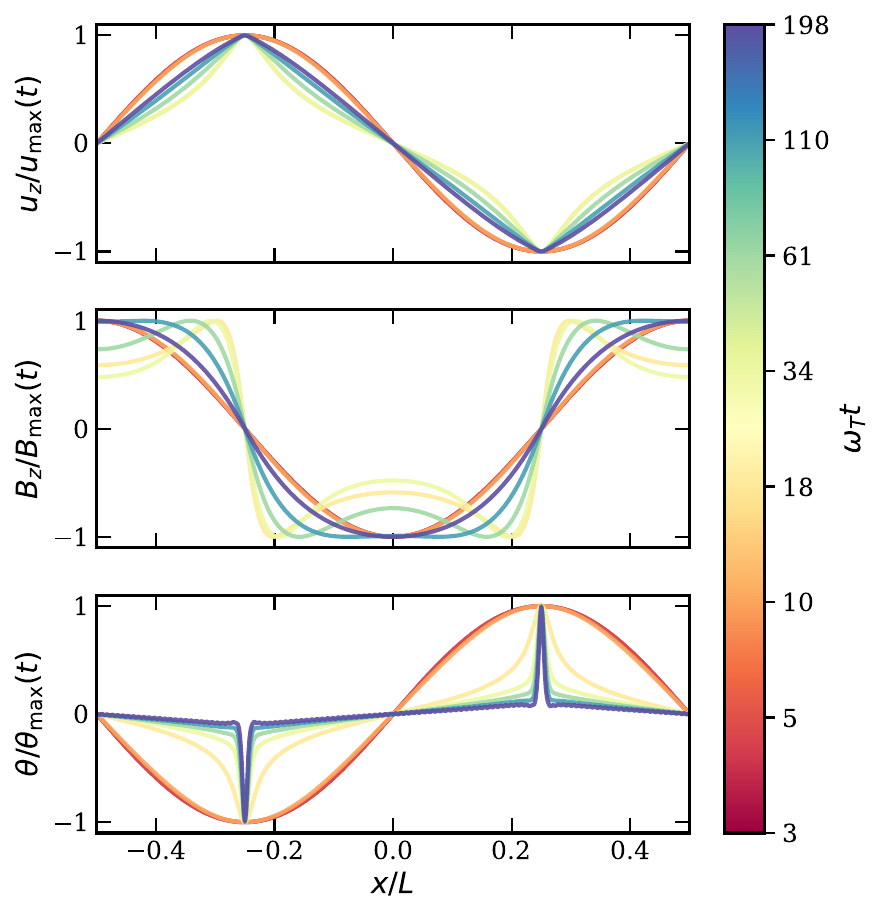}
	\caption{Horizontal profiles of $u_z$, $B_z$, and $\theta$ at different times of the 1D simulation shown in Fig.~\ref{fig:kin_pot_mag_energy}, normalized by their instantaneous maximum value. The phase of MTI exponential growth ends around $\omega_T t=10$, and the solution latches on an algebraically-growing elevator mode.}
	\label{fig:horizontal_profiles}
\end{figure}

\subsection{Numerical illustration}

In order to illustrate these ideas, especially the handover from the linear adiabatic MTI to the elevator solution, 
we run 1D {\sc SNOOPY} simulations of an MTI eigenmode, which, though idealised, produce physics that survives in our later 3D runs. We adopt parameters $N=0$, $\Pe = 100$, $\mathrm{Pr} = 0.01$, $\mathrm{Pm} = 1$, and a weak background magnetic field $\bm B_0 = 10^{-4} \bm e_x$ (see Section~\ref{sec:numerical-experiments} for further details on the numerical setup). The simulations use a resolution of 512 Fourier modes in the horizontal direction.
We show in Fig.~\ref{fig:kin_pot_mag_energy} the resulting time evolution of the kinetic, internal, and vertical magnetic energies.

The first stage of the evolution (up to $\omega_T t \sim 10$) corresponds to the MTI's linear phase, which agrees well with exponential growth at a rate $\sigma$ computed from the MTI dispersion relation, Eq.~(24) in \citetalias{Perrone2022}. At the end of this stage, rather than saturating, the perturbations continue growing algebraically. In particular, $u_{\mathrm{rms}} \propto t$, as predicted for an elevator flow, and indeed during this phase $B_z$ is much greater than the background field, which allows the handover from MTI mode to adiabatic elevator mode. The background field, however, is not completely subdominant during this phase and leads to growth in $\theta_{\mathrm{rms}}$ and $B_{\mathrm{rms}}$ (owing to the increase in the vertical component of the magnetic field) not included in Sections~\ref{sec:1d-model-equations} and \ref{sec:1d-solution}. 

Figure~\ref{fig:horizontal_profiles} plots the (normalised) spatial profiles of the solution at different times. By the end of the linear phase of growth, the mode structure begins to approximate the velocity triangle wave and magnetic square wave profiles discussed in \citetalias{Perrone2022}. The buoyancy variable, on the other hand, is peaked around the internal boundary layer encompassing the magnetic nulls. These profiles might be thought to serve as the `initial conditions' to the problem outlined in Section~\ref{sec:1d-model-equations}. As such, the two buoyancy peaks drive the further linear growth of $u$ and, helpfully, coincide with the extrema of its profile at this stage. As time progresses the vertical shear evident in the elevator intensifies, and (in more than on dimension) we expect it to break down due to Kelvin-Helmholtz instability.

\section{Numerical experiments}\label{sec:numerical-experiments}

We now conduct 3D numerical simulations of the adiabatic MTI, exploring the various ideas raised so far, namely: the emergence of large-scale MTI plumes, and their relation to the idealised elevator flows calculated in Section~\ref{sec:1d-model}; the long-time evolution of the elevators and their destabilisation by Kelvin-Helmholtz instability; the saturated values of the resulting turbulent kinetic energy, and its behaviour as parameters vary.

\subsection{Methods}

For our numerical simulations we employ a modified version of the pseudo-spectral code {\sc SNOOPY} \citep{Lesur2015}, where we included anisotropic heat conduction with a super time-stepping (STS) algorithm \citep{Alexiades1996}. In the {\sc SNOOPY} code, a Fourier transform is applied to Eqs.~\eqref{eq:div_eq}--\eqref{eq:buoyancy_eq}, and the Fourier coefficients are then advanced in time with a three-step Runge-Kutta algorithm. Nonlinear terms are computed in real space, applying a standard $2/3$ de-aliasing technique. Further details on the implementation of the STS can be found in \citetalias{Perrone2022}.

Our simulations use a resolution of either $128^3$ or $256^3$ Fourier modes (or grid points) in triply-periodic domains with an aspect ratio of 1:1:1. They are typically seeded with low-amplitude noise of $10^{-5}-10^{-4}$ (in code units) in the velocity field. The simulation domain is threaded initially with a horizontal field, of either net flux (NF) or zero-net flux (ZNF), i.e. $\bm B_0 = B_0 \bm e_y$ or $\bm B_0 = \sqrt{2} B_0  \sin(2 \pi z / L) \bm e_y$, for $B_0= 10^{-5}$ or $10^{-4}$, so that both configurations have the same initial magnetic energy. It is worth noting that configurations with net flux and strong initial magnetic fields are characterised by very different evolution and saturation compared to that of the MTI in the weak-field limit \citepalias[described by the scaling laws in Section~\ref{sec:mti-saturation-adiabatic}; see][]{Perrone2022,Perrone2022a}. For the Peclet numbers explored in this work, however, initial amplitudes of $B_0= 10^{-5}$ or $10^{-4}$ are well within the weak-field limit.

\begin{figure}
    \centering
    \includegraphics[width=1.0\linewidth]{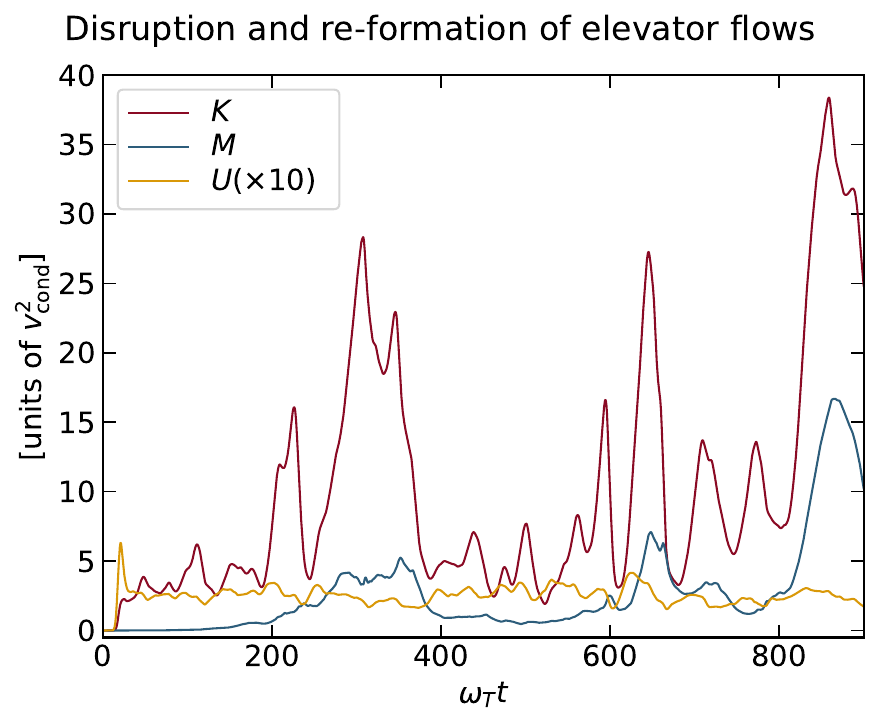}
	\caption{Evolution of kinetic, magnetic, and thermal energy in the representative adiabatic MTI run described in Section~\ref{sec:cycles-elevator}. The kinetic energy undergoes intermittent bursts that reflect runaway elevator flows.  }
	\label{fig:energies_disruption_formation_KHI_lowres}
\end{figure}

\begin{figure*}
    \centering
    \includegraphics[width=0.9\linewidth]{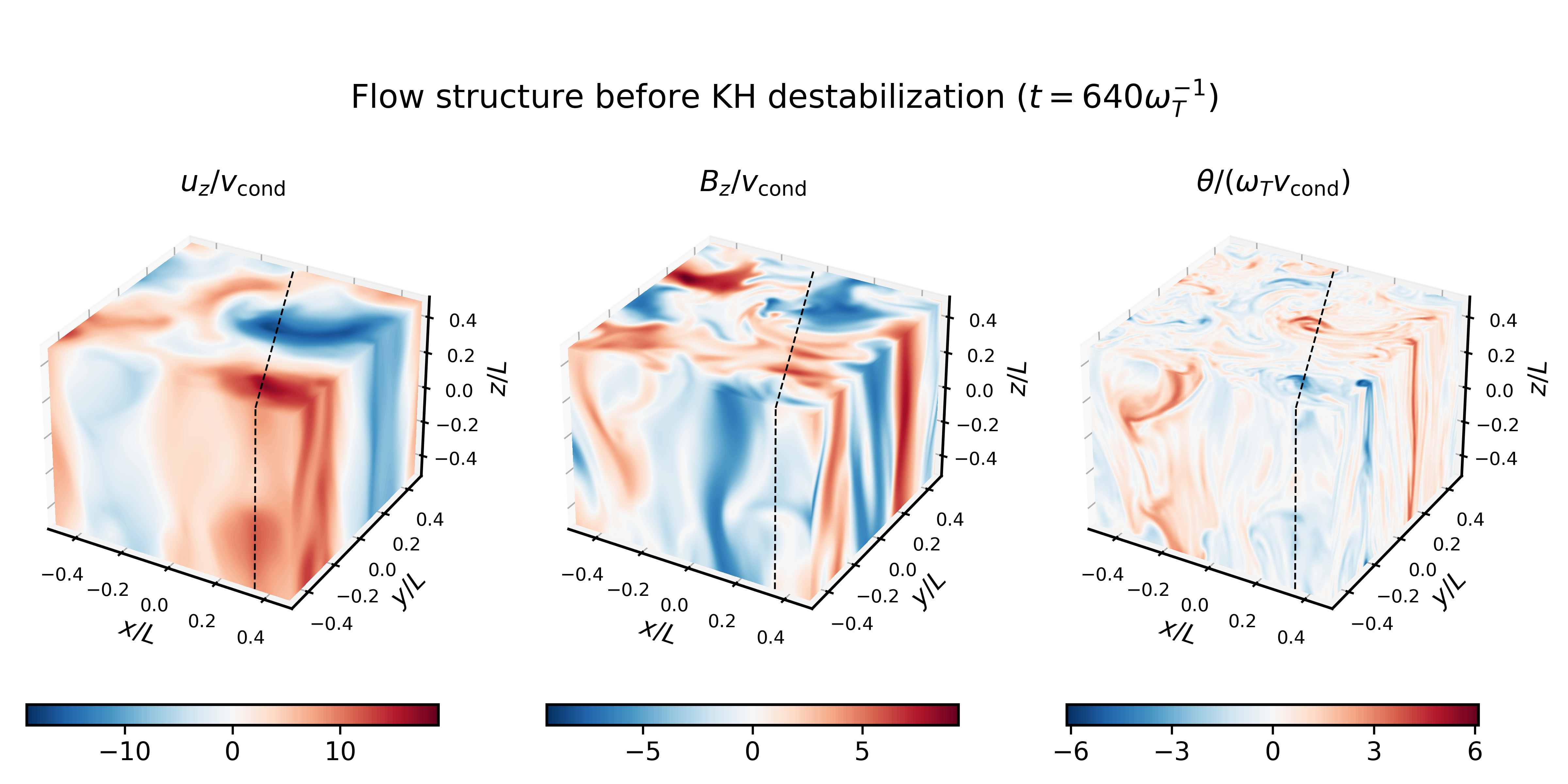}
	\caption{Snapshots of vertical velocity, vertical magnetic field and the buoyancy variable at a moment near the peak of one of the `elevator bursts' shown in Fig.~\ref{fig:energies_disruption_formation_KHI_lowres}, at $\omega_T t= 640$. Parameters of the run are given in Section \ref{sec:cycles-elevator}. The black dashed line indicates the vertical cut employed in Fig.~\ref{fig:transv_long_cut_avg}.}
	\label{fig:flow_structure_before_KHI_lowres}
\end{figure*}

\begin{figure}
    \centering
    \includegraphics[width=0.9\linewidth]{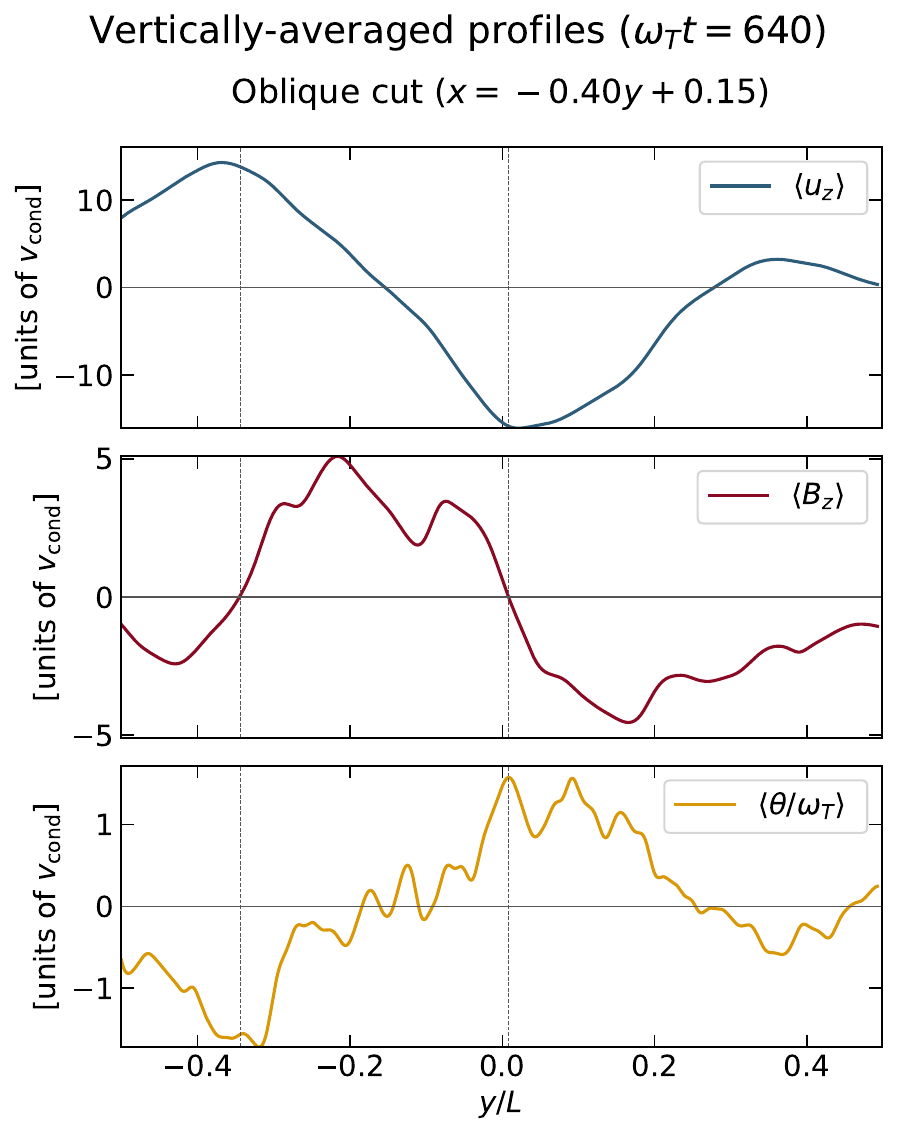}
	\caption{One-dimensional, vertically-averaged profiles of vertical velocity and magnetic field, and the buoyancy variable along the vertical cut shown in Fig.~\ref{fig:flow_structure_before_KHI_lowres}. Two vertical dashed lines are drawn at the locations of the nulls of $\langle B_z \rangle$.}
	\label{fig:transv_long_cut_avg}
\end{figure}

\begin{figure}
    \centering
    \includegraphics[width=0.9\linewidth]{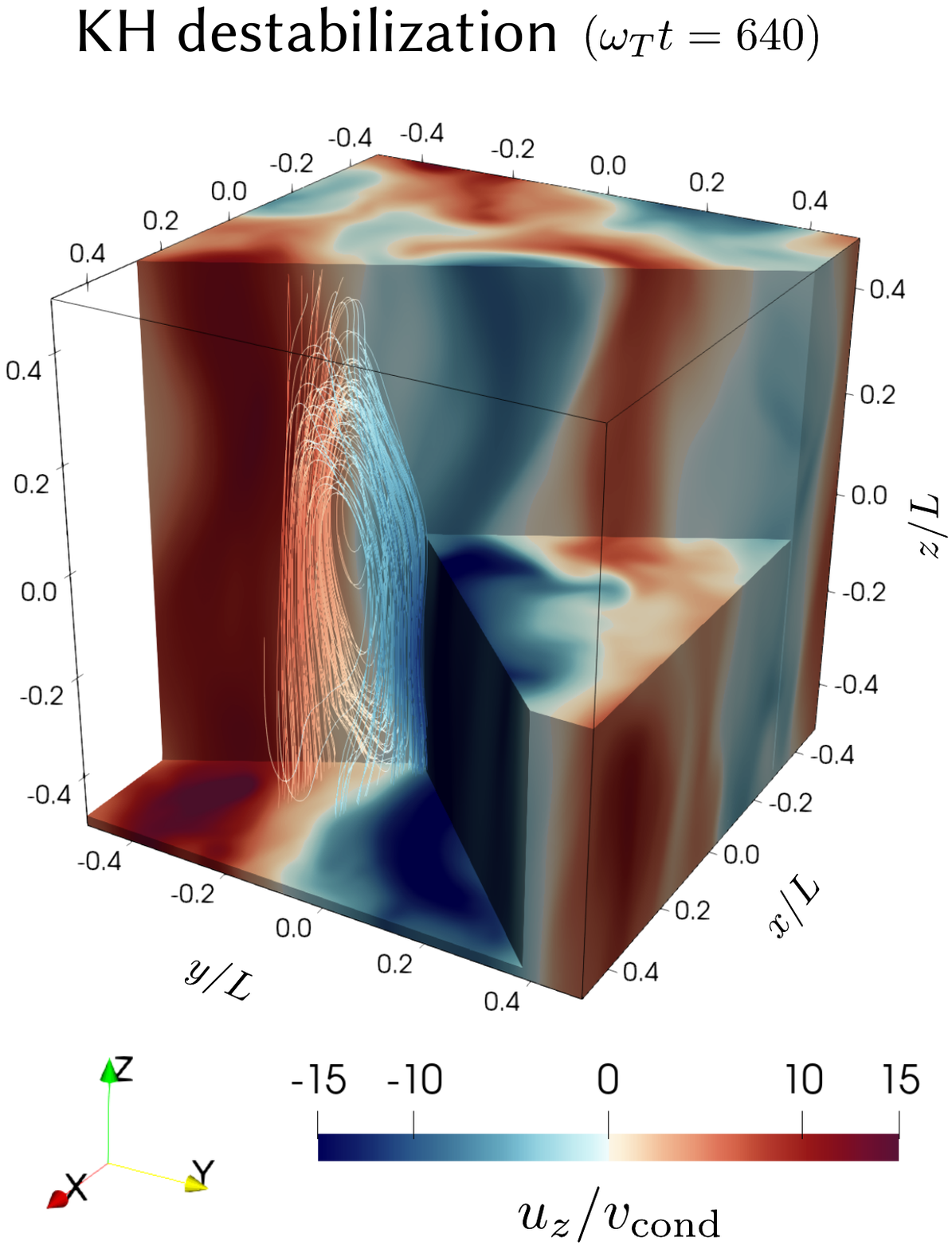}
	\caption{A rendering of the flow structure during destabilization by KHI in the run described in Sections \ref{sec:cycles-elevator} and \ref{sec:khi-regulation}. We plot the vertical velocity and superimpose the streamlines along the cut described by the dashed line in Fig.~\ref{fig:flow_structure_before_KHI_lowres}. }
	\label{fig:KH_post_uz_ticks}
\end{figure}

\subsection{Cycles of elevator flows}\label{sec:cycles-elevator}

To set the scene, we show results of a representative 3D run, exemplifying the main dynamical ingredients when $N=0$. Here $\Pe = 800$, $\mathrm{Pr} = 0.25$, $\mathrm{Pm} = 4$, and we impose a weak horizontal field of $ B_0 = 10^{-4}$.  We choose a lower resolution here, $(128)^3$, so as to simulate many cycles of elevator growth and destruction.

In Fig.~\ref{fig:energies_disruption_formation_KHI_lowres} we plot the box-integrated energies versus time. After an initial linear phase,  the system enters a disordered state characterised by a base-level of turbulent kinetic energy $K \gtrsim \varv_{\mathrm{cond}}^2 \equiv \ell_{\chi}^2\omega_T^2$, punctuated by strong bursts of magnitude (10--30) $\varv_{\mathrm{cond}}^2$. Notably, the kinetic energy is dominated by the vertical component during the bursts, and accompanying variations in magnetic and thermal energies are not nearly so strong or well correlated. 

The bursts correspond to relatively large-scale vertical flows, which we identify as relations of the 1D elevator flows discussed in Section~\ref{sec:1d-model}. To reveal their structure, we plot the 3D vertical flow and magnetic field profiles in Fig.~\ref{fig:flow_structure_before_KHI_lowres} at a burst peak, immediately preceding breakdown. The left panel shows that the domain clearly splits into regions hosting large-scale up-drafts and down-drafts, superimposed on significant smaller-scale disorder. To better confirm the identification of this structure as an elevator flow, Fig.~\ref{fig:transv_long_cut_avg} plots the 1D profiles of selected variables along the dashed cut-line in Fig.~\ref{fig:flow_structure_before_KHI_lowres}, which we compare with Fig.~\ref{fig:horizontal_profiles}. The vertical velocity $u_z$ exhibits a single period of a near-triangular wave, while the vertical magnetic field possesses zeros at precisely the locations of the maxima of $u_z$, and the extrema of $\theta$ are in anti-phase with $u_z$. These features agree well with the 1D solutions of Section~\ref{sec:1d-model}, and thus firm up the identification.

\begin{figure*}
    \centering
    \includegraphics[width=0.9\linewidth]{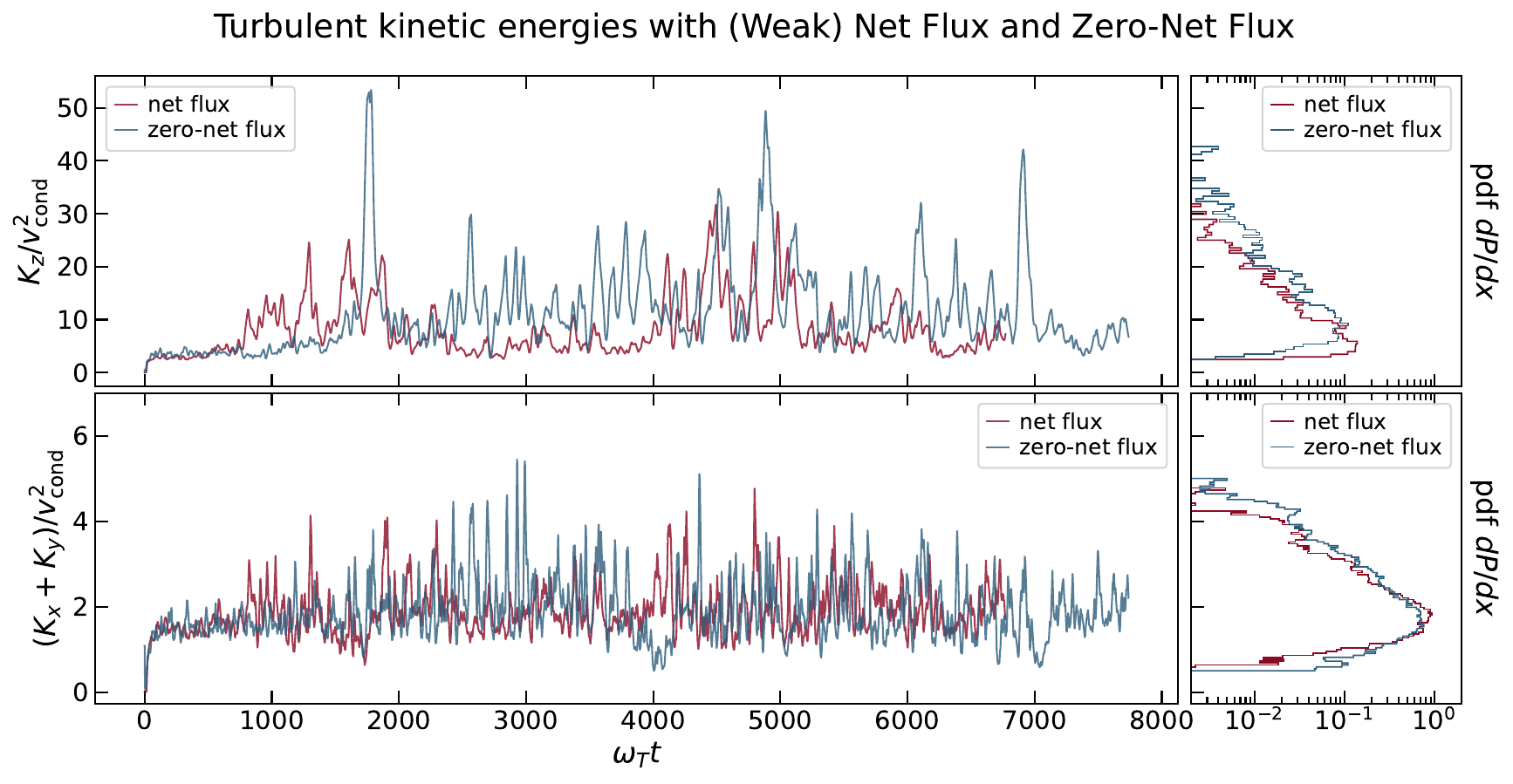}
	\caption{Long timeseries of the vertical kinetic energy (top left panel) and the horizontal kinetic energy (lower left panel) for a net flux and ZNF run, as described in Section~\ref{sec:parameter_survey}. In addition, we plot the probability distribution functions (pdf) of their fluctuations in the right panels. }
	\label{fig:NNF_weak_NF_compare_timeseries}
\end{figure*}

\begin{figure}
    \centering
    \includegraphics[width=0.9\linewidth]{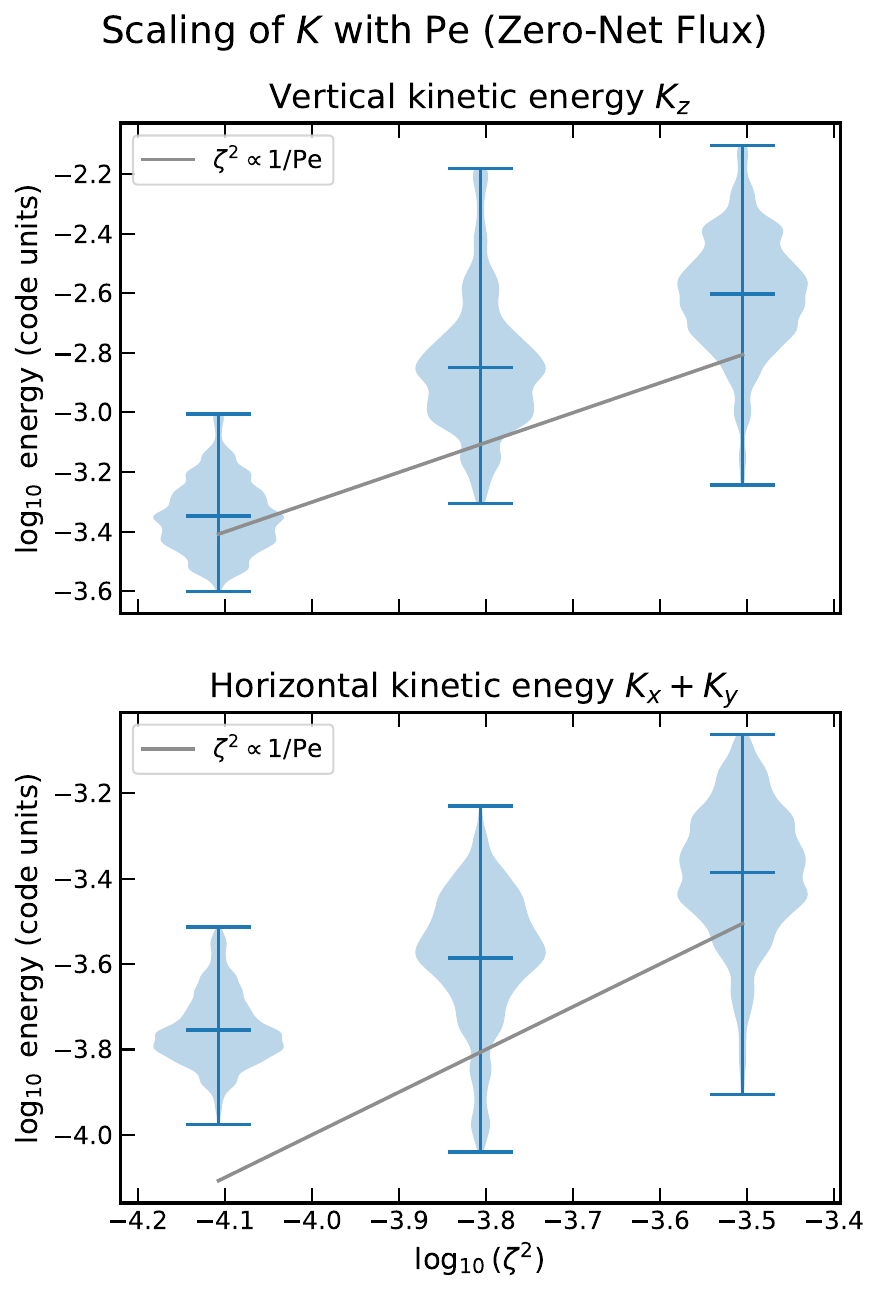}
	\caption{Violin plots for ZNF runs that describe the scaling of the vertical and horizontal kinetic energies with Pe. For other parameters and discussion, see Section~\ref{sec:parameter_survey}.}
	\label{fig:NNF_Pe_compare_scaling_KE_violin}
\end{figure}

\subsection{Regulation of adiabatic MTI by parasitic KHI}\label{sec:khi-regulation}

As predicted in Section~\ref{sec:mti-saturation-adiabatic}, the strong shear that develops in these (disordered) elevator flows makes them vulnerable to attack by KHI. These ultimately break down the large-scale flow structure, though do not always completely destroy it. In Fig.~\ref{fig:KH_post_uz_ticks} we plot a 3D rendering of $u_z$ in colour, along the oblique cut described by the dashed line in Fig.~\ref{fig:flow_structure_before_KHI_lowres}. Superimposed are the velocity streamlines across the cut, also coloured by $u_z / \varv_{\mathrm{cond}}$. As is clear from the large-scale roll, the KHI has developed upon the strong shear, leading to the disintegration of the large-scale flow.

Sometimes immediately after breakdown, an elevator flow begins growing again, which is often the case when the preceding flow is not fully destroyed. In particular, an imprint of the large-scale horizontal variation of $\theta$ remains and accelerates the elevator flow once again (see Eq.~\ref{eq:1Deqns}a). This is the case, for example, for the two consecutive peaks at $\omega_T t = 300$ and $\omega_T t = 350$ in Fig.~\ref{fig:energies_disruption_formation_KHI_lowres}.
In other instances, some period of time needs to pass, during which the flow remains generally disordered, before the advent of another strong elevator flow. We speculate that in these cases, the system must wait for the stochastic generation of a large-scale sinusoidal $\theta$ in the background turbulence, which can then rapidly accelerate the next elevator.

\subsection{Parameter survey and scaling laws}\label{sec:parameter_survey}

In this section we present results for several other parameter configurations so as to deduce general trends. The burstiness of the underlying dynamics necessitates very long time series (with averages undertaken over >1000 time units). Figure~\ref{fig:NNF_weak_NF_compare_timeseries} shows two such example long-term runs for a ZNF and a weak net flux set-up ($\bm B_0 = 10^{-5} \bm e_y$). In both cases, $\Pe=6400$, $\mathrm{Re} = 19200$, $\mathrm{Pm} = 2$, and the resolution is $(256)^3$. In line with our fiducial run in Section \ref{sec:cycles-elevator}, on longer times the kinetic energies of both runs settle on a base level of $\gtrsim \varv_{\mathrm{cond}}^2$, with intermittent bursts concentrated in vertical motion. 
A similar situation applies to the magnetic fields, which in our ZNF run are kept at approximate equipartition with the turbulent velocities by a small-scale dynamo, and that also exhibit an intermittent behavior with dominant vertical amplitudes.

The ZNF run, in fact, exhibits elevated energy levels and more frequent and violent bursts. This is quantified by the the probability distribution functions (pdf) plotted in the right panels of Figure~\ref{fig:NNF_weak_NF_compare_timeseries}. For the vertical velocity field, the ZNF pdf has a significantly longer tail at higher energies, and is slightly shifted generally to higher energies. On the other hand, the horizontal kinetic energy pdfs are similar. 
It is unclear whether the reduced levels of turbulence are due to insufficient sampling and statistical variability, or rather due to the different magnetic field topology. One possible explanation is that the presence of a horizontal net magnetic field (no matter how weak, since anisotropic conduction only depends on $\bm b = \bm B / B$) allows the temperature fluctuations to diffuse more effectively in the horizontal plane, smoothing out over time any coherent density blobs and thus reducing the vertical acceleration of the flow.

Informed by our discussion in Section~\ref{sec:mti-saturation-adiabatic}, we next vary the conduction length (i.e., Pe) so as to obtain a scaling law for the saturated kinetic energy, and thereby check whether the KHI regulates saturation in the way predicted ($K\sim \ell_{\chi}^2 \omega_T^2$ in the Boussinesq regime). One difficulty here is that our effective range of Pe is relatively limited. Raising Pe, while keeping Pr constant, makes the viscous scales difficult to resolve; on the other hand, small Pe, for a given Pr, leads to very powerful bursts, because the KHI is increasingly suppressed by viscosity, and requires shorter integration timesteps. We thus only examined $\Pe = 3200, 6400, 12800$, which translates to $\mathrm{Pr} = 0.167, 0.333, 0.667$ keeping the viscosity constant. A ZNF configuration was adopted in each case.

Figure~\ref{fig:NNF_Pe_compare_scaling_KE_violin} presents our results using `violin plots', which show not only the average vertical and horizontal $K$, but also its probability distribution: the middle horizontal bar is the average, while the upper and lower bars are the maximum and minimum values achieved. As is clear, despite the burstiness, the system does adhere to a scaling law, which in the case of vertical kinetic energy is consistent with a $\Pe^{-1}$ proportionality in code units. The magnetic energies also roughly follow the kinetic energy due to a small-scale dynamo in both our $\Pe=3200$ and $\Pe=12800$ ZNF runs. In real units, this transforms to $K\sim \ell_{\chi}^2 \omega_T^2 $, as predicted in Section~\ref{sec:mti-saturation-adiabatic}, at least for the large Pe regime we are testing here.  The implication is that  KHI is sufficiently efficient to saturate the MTI without recourse to box effects, at least to leading order in our runs. Plausibly, when $\ell_{\chi}$ gets longer (Pe smaller), then box effects re-emerge.

\section{Discussion}\label{sec:discussion}

In this section, we apply to observed ICM data the scalings predicted heuristically in Section~\ref{sec:mti-saturation-adiabatic}, and confirmed numerically in Section~\ref{sec:numerical-experiments}. Our starting point is the following Spitzer formula for thermal diffusivity \citep[e.g.,][]{Kunz2012}:
\begin{align}
	\chi_\text{S} = 4.98 \times 10^{32} \left( \frac{k_\mathrm{B} T}{5 ~\si{keV}}\right)^{5/2} \left( \frac{n}{10^{-4}~\si{cm^{-3}}}\right)^{-1} \si{cm^2.s^{-1}},
\end{align}
where $T$ is the electron temperature and $n$ the number density. It is important to stress that, in weakly-collisional plasmas like the ICM, plasma kinetic instabilities excited at the ion and electron Larmor radii can act a source of enhanced scattering for the heat-carrying thermal electrons, thereby reducing their mean-free-path and suppressing the effective thermal diffusivity below Spitzer levels. We model suppression of thermal conductivity by kinetic micro-instabilities and tangled magnetic fields by a linear suppression factor $f < 1$ and write $\chi = f \chi_\mathrm{S}$ \citep[for limitations of this parametrization in the case of the MTI and whistler instability see][]{Perrone2024a,Perrone2024}. Kinetic studies \citep{Roberg-Clark2016,Komarov2018,Yerger2025} and observational estimates \citep{Vikhlinin2002,Bale2013,Richard-Laferriere2023} typically constrain $f$ between $f = 0.01 - 0.1$.

Next, we take a model cluster and assume that its temperature has a radial dependence at large radii as $T(r) = T_0 (r/R_0)^{-a}$, where $T_0$, $R_0$ are reference values, and $a>0$ is the slope. Assuming hydrostatic equilibrium, we can recast the MTI frequency $\omega_T$ (Eq.~\ref{eq:freq}) as a function of radius as $\omega_T^2 (r) = a G M_< (r) /r^3$, where $G$ is the gravitational constant, and $M_< (r)$ is the enclosed mass within a sphere of radius $r$. Choosing reference values for massive clusters and $a=0.4$ \citep[extrapolated from][]{Ghirardini2019}, we obtain:
\begin{align}
    \omega_T = 4.25 \times 10^{-17} \left( \frac{a}{0.4} \right)^{1/2} \left( \frac{M}{10^{15} M_\odot} \right)^{1/2} \left( \frac{r}{\si{Mpc}} \right)^{-3/2} \si{s^{-1}}, \label{eq:omega_T_estimate}
\end{align}
which corresponds to timescales of $1/\omega_T \sim 750 ~\si{Myr}$. 

Noting the heuristic arguments culminating in the lower bound in Eq.~\eqref{eq:range_mti_saturation} and, in particular, the simulation data represented in Fig.~\ref{fig:NNF_weak_NF_compare_timeseries}, we estimate a turbulent kinetic energy at saturation of $K \sim 10 ~\varv_{\mathrm{cond}}^2 = 10 ~ \ell_\chi^2 \omega_T^2$. Taking into account the $(\gamma - 1)/\gamma$ rescaling of thermal diffusivity in the Boussinesq equations (Eqs.~\ref{eq:div_eq}--\ref{eq:buoyancy_eq}), this corresponds to a turbulent rms velocity of $u_\mathrm{rms} \approx 2.83 f^{1/2} (\chi_\mathrm{S} \omega_T)^{1/2}$.
Choosing values appropriate for the periphery of massive galaxy clusters, and using Eq.~\eqref{eq:omega_T_estimate}, we find typical turbulent rms velocities of:
\begin{align}
    u_\mathrm{rms} \simeq 1300 ~ &\si{km.s^{-1}} \left( \frac{a}{0.4} \right)^{1/4} \left( \frac{f}{0.1} \right)^{1/2} \left( \frac{k_\mathrm{B} T}{5 ~\si{keV}}\right)^{5/4}  \nonumber \\
    \times &\left( \frac{n}{10^{-4}~\si{cm^{-3}}}\right)^{-1/2} \left( \frac{M}{10^{15} ~M_\odot} \right)^{1/4} \left( \frac{r}{\si{Mpc}} \right)^{-3/4}.
\end{align}
This estimate is substantially larger than that for the stably stratified MTI, examined in \citetalias{Perrone2022a}, where we estimated $u_\mathrm{rms} \simeq 130 ~ \si{km.s^{-1}}$ using similar reference values for the cluster periphery. Indeed, it is comparable to the adiabatic sound speed, $c_{\mathrm{s}} = (\gamma k_\mathrm{B}T / \mu m_{\mathrm{u}})^{1/2}\simeq 1200~\si{km.s^{-1}}$ for $k_\mathrm{B}T = 5 ~\si{keV}$, and mean molecular weight $\mu \simeq 0.588$, corresponding to fully ionized primordial hydrogen-helium gas ($m_{\mathrm{u}}$ is the atomic mass unit), in agreement with Eq.~\eqref{estimate}. While it is true that at such speeds the Boussinesq approximation breaks down, the dimensional argument supporting Eq.~\eqref{estimate} is more general. Furthermore, the example of convection in the Solar surface, which can ultimately achieve supersonic velocities \citep[e.g.,][]{Nordlund}, demonstrates that compressibility itself need not always impose a cap on the speed of buoyant plumes.

Finally, it is worth contrasting our estimates with previous numerical work, in semi-global or fully global geometries \citep{McCourt2011, Parrish2012c, McCourt2013a}, which also suggest that the MTI could provide significant turbulent pressure support. However, it is likely that the turbulent speeds measured in these earlier simulations were boosted by unrealistically  large (unsuppressed) thermal conductivities and temperature gradients, or an unstable entropy gradient in the case of \citet{McCourt2011}. Our work emphasises the importance of a flat entropy profile in permitting energetic large-scale plumes, even when the thermal conductivity is (relatively) suppressed.

\section{Conclusion}\label{sec:conclusions}

In the far outskirts of galaxy clusters, a number of observations suggest that the radial entropy profile may be shallower than predicted by models of gravitational accretion \citep[][]{George2009,Simionescu2011,Urban2011,Walker2012,Walker2013,Fusco-Femiano2014,Ghirardini2017,Simionescu2017}. Given that a stable entropy gradient is an essential ingredient in magnetothermal turbulence, we expect the MTI's behaviour to differ significantly from theoretical predictions in these outermost regions \citepalias{Perrone2022}. In this paper we present a mix of heuristic arguments, semi-analytical calculations, and local simulations (in the Boussinesq approximation) to characterise the MTI saturation when the entropy is constant. 

We argue that, in the absence of buoyancy, MTI plumes will develop unimpeded until they become vulnerable to Kelvin-Helmholtz instability (KHI). The MTI's saturation then follows a pattern of bursts comprising the development of runaway large-scale plumes and their subsequent breakdown in shear turbulence. The average kinetic energy of this state we estimate as $\sim \varv_{\mathrm{cond}}^2= \ell_{\chi}^2\omega_T^2$, though with strong deviations (during bursts) an order of magnitude larger. Thus, in principle, even when buoyancy is absent, the MTI can saturate without calling on compressibility, shocks, or the global structure of the cluster. That said, depending on the amplitude of $\varv_{\mathrm{cond}}$ \citep[and, in particular, its potential suppression by microturbulent processes;][]{Roberg-Clark2016,Roberg-Clark2018,Komarov2018,Drake2021,Perrone2024,Perrone2024a,Yerger2025,Choudhury2025}, the MTI plumes could reach the scale-height of the cluster and potentially shock. Indeed, for sufficiently large thermal diffusivity, energetic MTI plumes could collectively provide significant non-thermal pressure support, as certain observations suggest \citep{Ettori2019,Eckert2019}. Even accounting for realistic levels of heat conduction suppression to 10\% of the Spitzer value we find that the resulting MTI-driven turbulent motions approach transonic velocities.

The simulations we present are local and utilise the Boussinesq approximation, which imposes potential limitations. For small $\chi$, this approximation is probably acceptable, but even then the largest bursts might be reaching a regime outside the domain of applicability, in which case vertically stratified \citep{Bogdanovic2009,Kunz2012} or global models \citep{Kempf2025,Kempf2025a}, would provide more secure descriptions of the MTI saturation. In particular, simulations in cosmological settings would allow us to study the interplay of the MTI with other sources of turbulence in the periphery of galaxy clusters, such as mergers and gravitational accretion \citep{Talbot2024}.

\section*{Acknowledgements}
The authors thank the anonymous reviewer for a set of helpful comments and David Hosking and Jean Kempf, who generously read through a previous version of the manuscript. LMP gratefully acknowledges support by the European Research Council under ERC-AdG grant PICOGAL-101019746.

\section*{Data Availability}

The data underlying this paper will be shared on reasonable request to the corresponding author.



\bibliographystyle{mnras}
\bibliography{MTI-Thesis-Bibliography} 



\appendix




\bsp	
\label{lastpage}
\end{document}